# New insight into kinetics behavor of the structural formation process in Agar gelation

Y.Z.Wang[1,2*], X.F.Zhang[2] and J.X.Zhang[1*]

[1]*State Key Laboratory of Optoelectronic Materials and Technologies, and School of Physics and Engineering, Sun Yat-sen University, Guangzhou, 510275, China*

[2]*Bioengineering Program and Department of Mechanical Engineering and Mechanics, Lehigh University, 19 Memorial Drive West, Bethlehem, Pennsylvania 18015, USA*

A time-resolved experimental study on the kinetics and relaxation of the structural formation process in gelling Agar-water solutions was carried out using our custom-built torsion resonator. The study was based on measurements of three naturally cooled solutions with agar concentrations of 0.75%, 1.0% and 2.0% w/w. It was found that the natural-cooling agar gelation process could be divided into three stages, sol stage (Stage I), gelation zone (Stage Ⅱ) and gel stage (Stage Ⅲ), based on the time/temperature evolutions of the structural development rate (SDR). An interesting fluctuant decaying behavior of SDR was observed in Stage Ⅱ and Ⅲ, indicative of a sum of multiple relaxation processes and well described by a multiple-order Gaussisn-like equation: $dG'/dt \equiv \sum_{n=0}^{m} dG'/dt^{(n)} = \sum_{n=0}^{m} K_n \exp[-2(T-T_n)^2/W_n^2]$. More interestingly, the temperature dependences of the fitted values of $W_n$ in Stage II and Stage III were found to follow the different Arrhenius laws, with different activation energies of $E_{aⅡ}$= 39-74 KJ/mol and $E_{aⅢ}$~7.0 KJ/mol. The two different Arrhenius-like behaviors respectively suggest that dispersions in Stage Ⅱ be attributed to the relaxation of the self-assembly of agar molecules or the growth of junction zones en route to gelation, in which the formation or fission of hydrogen bonding interactions plays an important role; and that dispersions in Stage Ⅲ be attributed to the relaxation dynamics of water released from various size domains close to the domain of the viscous flow of water during the syneresis process.

---

[*]To whom correspondence should be addressed. E-mail: wangyizhen80@gmail.com (Y.Z.Wang) or stszjx@mail.sysu.edu.cn (J.X.Zhang).

## 1. Introduction

Biopolymer gels with three-dimensional fibrous network structures constitute a special class of soft matter. They are widely used in many industrial fields[1-3], such as chromatography and filtration, photography, hematology, food technology, and pharmacy. Their unique mechanical and rheological properties, as well as the peculiar gelling/melting processes, have attracted considerable scientific interest connected to the study of phase transitions, scaling, and criticality and universality[1-3]. The study of these materials also adds to the understanding of the fundamental physics of the sol-gel/gel-sol transition[3-7]. Despite many experimental and theoretical efforts[3-21] from scientists in the past, the achieved progress is far from satisfactory. The formation kinetics of the network structures in biopolymers and their associated gelation mechanism still remain poorly understood[6-7,12,15-18,21].

Agar is a gel-forming polysaccharide extracted from seaweeds with a sugar skeleton consisting of alternating 1,3-linked β-D-galactopyranose and 1,4-linked 3,6 anhydro-α-L-galactopyranose units; it is used as a model biopolymer in gelation[2]. Time-resolved experiments combining techniques such as rheology, scattering, and measurements of turbidity by spectrometer or differential scanning calorimeter are ideal for investigating the agar gelation mechanism. In the past 30 years, many time-resolved experiments[4,6-10,13-27] have been performed to extensively study the micro-structural, mechanical and rheological properties of agar or agarose gels; however, some resulting conclusions seem inconsistent. For example, based on turbidity and rheology measurements, X.Y. Liu's group[15] recently suggested that when a quench rate as large as 30 deg C/min is applied, agarose gelation is initiated by a nucleation and growth mechanism rather than the previously suggested mechanism of spinodal decomposition. More recently, based on the results of light scattering and rheology experiments, Bohidar's group[16] suggested that agar gelation is a complex, gradual process, and the self-assembly of agar molecules en route to gelation indicated the simultaneous growth of self-similar aggregates and larger macro-domains due to spinodal demixing. Clearly, most of the previous inconsistencies are due to insufficient experimental investigation of the kinetics and equilibrium processes of the sol-gel/gel-sol transitions of agar or agarose gels, as well as the effect of the gelation conditions on the gels' microstructure and viscoelastic properties[15-18,22-27].

In the present work, we report a time-resolved experiment based on custom-developed torsion resonator[28-30], with which we examine the gelation behavior of three aqueous agar solutions with agar concentrations ($C$) of 0.75%, 1.0% and 2.0% w/w during natural cooling from 80$^\circ$C to room temperature (~25$^\circ$C). The time ($t$) evolution of the temperature ($T$) and the mechanical properties ($tg\phi$, $G'$ and $G''$, and $dG'/dt$) of these agar-water gel systems were investigated. Our primary focus was the influence of temperature on the structure development rate (SDR) or gelation rate, defined by $dG'/dt$, of the agar sol-gel transition during natural cooling. Our experimental results show that the agar sol-gel transition during natural cooling is a gradual process that exhibits a complex gelation mechanism involving multiple relaxation processes, which can be characterized by multiple-order Gaussian-like kinetics of the fluctuant temperature-dependent $dG'/dt$: $dG'/dt = \sum_{n=0}^{m} K_n \exp[-2(T-T_n)^2/W_n^2]$. The temperature ($T_n$) dependences of the obtained $W_n$ (or $K_n$) at high temperature ($>T_{pe}$, during the agar sol-gel process) and at low temperature ($<T_{pe}$, during the following syneresis process) were typically found to show different Arrhenius behaviors, yielding distinct apparent activation energies. These findings might provide a significant insight into the mechanical properties and distinctive structure development rates of agar sols that dynamically and naturally evolve to form gels.

## 2. Experimental section

Agar powder with a molecular weight of ~120kDa was supplied by *Guangdong Huankai Microbial Sci. & Tech. Co. Ltd*. The aqueous agar solutions with concentrations of 0.75%, 1.0% and 2.0% w/w were prepared by dissolving the agar powder in deionized water at 100°C. After 30min of boiling, the solution was filtered at 80-90°C through 0.22μm filters and then quickly loaded into the preheated (80°C) sample cell (Fig.1.c-1) of our torsion resonator apparatus[28-30]. The sample-loaded cell, made from Teflon material and of the annular-cup type[29a-d], was contained in a water-controlled temperature housing (Fig.1.c-5) with the temperature stability better than ±0.01°C and connected to a HS-4 standard thermostatic water-bath[31a] (Fig.1.a). Temperature control was accomplished by circulating an ethylene glycol-water mixture from the constant-temperature water bath through passages machined in the housing[29a-c].

After preparation at 80°C, the loaded hot agar sol was subsequently allowed to slowly cool from 80°C to room temperature (~25°C) by switching off the power supply (or power button, see Fig.1.a) of the constant temperature water-bath and letting the ethylene glycol-water mixture in the temperature housing naturally cool to room temperature, as in previous studies[1-3,7-12,22-24].

During the cooling process, the temperature of the agar-water gel system was measured with a high-precision thermocouple (Fig.1.c-3) carefully inserted into the sample fluid. Changes of ±0.05$^o$C can be readily detected using a high-precision *Yudian UGU* microprocessor temperature controller[31b] (Fig.1.b).

Simultaneously, the time dependence of the mechanical and viscoelastic properties of the agar-water gel system were detected by our torsion resonator[29], which works in the constant applied-stress mode and at a frequency of *10Hz*. This modified torsion resonator, with a phase angle measurement uncertainty of less than *±0.0003$^o$* and a displacement signal measurement uncertainty of better than *1%*, has already been successfully used to gain insights into the viscoelasticity of aqueous glycerol solutions[29a], polyethylene oxide solutions[29c], semi-dilute polyacrylamide-water solutions[29b], and granular materials[29d], as well as the water-to-ice transition[29e]. In the reported time-resolved experiments, narrowing the gap between the resonator and the sample cell to *~2.0mm* allowed the torsion resonator apparatus to work at the gap-loading condition, similar to a commercial stress-controlled dynamic rheometer[30]. To ensure that all the measurements were made in the linear viscoelastic regime, and to avoid destruction of the structure of the agar gel and slippage of the gel during measurements, typical working amplitude of oscillations of the torsion resonator were restricted to be of the order of *1.0μm* or less by applying the slight constant external stress torque (τ) of *193μNm*.[29a-c]

Additionally, the sample cell and temperature housing of the torsion resonator were specifically designed to significantly reduce the temperature gradients in measured samples and to ensure thermal quasi-equilibrium in the sample during each measurement. As shown in Fig.1, the sample cell was designed with a small wall-thickness of less than 0.5 mm and a smaller gap of ~4.0 mm between the inner and outer walls. The temperature housing was made from red copper (1.0 mm thick) for good thermal conductivity and tightly wrapped in a Teflon jacket (Fig.1c-6) to reduce heat leakage; the sample cell was located just below 2/3 the height of the temperature housing to obtain a more stable and even temperature distribution in the measured samples. For a reference sample of pure water, this design allowed the temperature gradient in a pure-water sample at room temperature to be controlled to within $0.1^oC$.

Each torsion-resonator measurement reported in this paper was repeated at least three times to verify reproducibility.

## 3. Results and Discussions

Typical gelation kinetics of the agar-water gel system with C=2.0% w/w are shown in Fig. 2. The agar sol was naturally cooled to ~$30^oC$ at a rate of approximately $0.1^oC$/min, as indicated in Fig.2a. Based on Fig.2a and 2b, it is clear that both the measured temperature ($T$) of the gelling agar-water sample and the response strain ($\gamma$) of the oscillating cup (inserted into the measured agar solution under the constant applied-stress torque $\tau=193\mu Nm$) decrease gradually and continuously with increasing time during the gelation process. Such gradual and continuous decreases of the temperature and the response strain are distinctly different from the time-dependent temperature and strain behavior in the freezing process of water[29e,32-33]. The water-to-ice freezing process usually shows an abrupt change at the freezing point, representative of a typical first-order phase transition or a nucleation and crystallization process[32-33]. Nevertheless, no abrupt change was observed in the $T(t)$ and $\gamma(t)$ curves in Fig.2a and 2b, indicating that the mechanism of the agar sol-gel process upon naturally cooling differs from that of the freezing process of water. Also shown in Fig.2b is the time dependence of the measured apparent mechanical loss tangent ($tg\phi$). In terms of the relationships[29-30] $G'=A\times(\tau/\gamma)cos\phi$ and $G''=A\times(\tau/\gamma)sin\phi$ (*Here $A\equiv(2\pi RH_0)^{-1}$ is the apparatus constant, and R is the radius of the oscillating inverted cup, $H_0$ is the height of the cup inserted into the measured sample*), we can obtain the time dependence of the corresponding storage ($G'$) and loss ($G''$) moduli, given in Fig.2c. Fig.2c shows that $G'$ of the agar-water systems undergoing natural gelation increased significantly with decreasing temperature and increasing time, and the increment of $G'$ was much larger than that of $G''$ at the later stage of gelation. These results are quite similar to previous results[7-9,14-16,18,22-26,34-40] obtained by using a commercial dynamic rheometer, demonstrating the validity of our torsion resonator experiments.

Our focus here is the time-derivative of $G'$ (i.e. $dG'/dt$), as the time and temperature dependence of $dG'/dt$ can provide a measurement of the rate of gelation, reflecting the network formation of the agar-water system during the gelation process. As in previous studies[7,15,18,23-26,37-28], the differential coefficient of the time-$G'$ curve in Fig. 2c can be directly calculated by using *the ORIGIN* software package[31c]. This calculated value allows us to obtain the structure development rate (SDR) or gelation rate[7,15,18,25,38] characterized by $dG'/dt$, as given in Fig.2d.

Three stages (indicated as I, II and III in Fig. 2d), which were quite similar to the previous divisions[15-16] based on light scattering experiments and on turbidity and rheological measurements respectively, could be identified. Stage I ($T \geq T_g$) corresponds to the sol state, where the temperature is higher than 52°C and both $G'$ and $dG'/dt$ are extremely low. In this stage, the agar-water samples could be considered an amorphous and fully ergodic phase, consisting of distinct-size particles having Gaussian chain (random coil) or disconnected helix character[16]. As the sample temperature approaches the gelation temperature ($T_g$), the occurrence of a coil-to-helix transition[9,15-16,20] is possible. Agar particles of Gaussian or associated helix character slowly increase to a certain size as they lose their Gaussian character, becoming interconnected and developing ordered rod-like conformations or fiber junctions. As soon as the rods are formed, agar gelation (or network formation) starts, leading to a steep increase in both $G'$ and $dG'/dt$ and a steep decrease in the responded strain ($\gamma$) at time $t_g \approx 50$min (or at $T_g \approx 52^oC$), as shown in Fig. 2c-d. In Stage II ($T_g > T \geq T_{pe}$), the agar gelation process proceeds. The rigidity of the agar-water gel system continues to increase with time; its gelation rate rapidly increases at first and then decays gradually before leveling off and fluctuating around zero in Stage III ($T_{pe} > T$). The boundary temperature $T_{pe}$ at time $t_{pe} \approx 120$ min (between Stages II and III) could be defined to be the temperature at which an agar gel network in dynamic equilibrium or a pseudo-equilibrium state[15] is completely formed. Below $T_{pe}$ in stage III, the rigidity of such a pseudogel continues to increase, possibly due to syneresis[15-16,39], despite the much slower structural development rate. Values of the stage-dividing temperatures ($T_g$ and $T_{pe}$) for the three agar-water systems, C=0.75%, 1.0% and 2.0% w/w, are given in Fig.3c. Clearly, a higher agar concentration results in higher values of both $T_g$ and $T_{pe}$, indicating that the division of kinetics stages by SDR is greatly dependent on the polysaccharide concentration.

Because the natural cooling rate in the reported time-resolved measurements (~0.1$^o$C/min) was very low, the agar-water gel system can be considered to have reached the quasi-equilibrium state corresponding to the measured-temperature at each point of the time-temperature curve. Therefore, using the measured time-dependent temperature and corresponding $dG'/dt$ data (e.g. those in Fig.2a and 2d), it is possible to obtain the temperature dependence of the SDR during the natural-cooling gelation process for the three gelling agar-water systems (C=0.75%, 1.0% and 2.0% w/w), as given in Fig.3b. Corresponding temperature-$G'$ and temperature-$tg\phi$ behaviors are shown in Fig.3a.

Interestingly, from Fig.3b we find that following the rapid increase during the initial stage of the agar gelling process, the SDR exhibits a fluctuant decaying behavior[a] in Stages II and III. This is quite different from the previously reported behavior[7,15] in which, following an initial rapid increase, the SDR gradually and smoothly decays until it reaches a smaller value that is almost independent of the ageing time or ageing temperature.

From previous studies of multiple relaxation behaviors in soft materials[2-6,11-20,22-26,35-40], we argue that the fluctuant decaying behavior of the SDR shown in Fig. 3b indicates the presence of multiple relaxation processes during the agar gelation process, each having a SDR with different magnitude and temperature dependence. Analogously, based on the assumptions that the total relaxation of the SDR should be the sum of the individual relaxation of SDR$^{(n)}$ (i.e. $dG'/dt^{(n)}$) and that each individual relaxation satisfies a Gaussian-like temperature dependence[2-4,7,20,22-26,35-39], a multiple order Gaussian equation for describing such temperature-SDR behavior is given by the following equation:

$$\frac{dG'}{dt} = \sum_{n=0}^{m} \frac{dG'^{(n)}}{dt} = \sum_{n=0}^{m} K_n e^{-2(\frac{T-T_n}{W_n})^2} \qquad (1)$$

where $K_n$ denotes the magnitude of each dispersion, $T_n$ is the temperature at the peak maximum and $W_n$ is the characteristic width of each $dG'/dt^{(n)} \sim T$ dispersion profile; $m$ characterizes the number of dispersions[41].

---

[a] Factually, similar fluctuant decaying behavior had ever been detected in previous experimental studiess[7,18,20,25,38], whereas it was unfortunately overlooked.

To quantitatively understand the properties of the relaxation behavior in Fig.3b, ORIGIN software[31c] was used to find a best nonlinear fit of Eq.1 to the experimental $dG'/dt^{(n)} \sim T$ data in Fig.3b. Values of the $K_n$, $T_n$ and $W_n$ parameters of each of the component dispersions used to fit the experimental data are given in Fig.4b and 4c. A typical fitting result for the $C=2.0\%$ w/w agar-water gel system is shown in Fig.4a. From Fig.4a, eight relaxation processes in Stages II and III (for the case of $C=2.0\%$ w/w), designated as n=0-7, were unambiguously identified from data in the studied temperature range. From previous studies[2-6,11-20,22-26,35-39], the $n=0$ dispersion with the maximum SDR magnitude ($K_{n=0}$) at $T=T_0$ might originate from the formation of junction zones in polymer-rich phases due to the assembly of the disconnected agar fibers or helices[9,15-16,25]. The second- and third-rate processes can be idealized as corresponding to the subsequent enlargement of junction zones and the local coagulation of polymer-rich phases. These processes lead the free energy of the whole system to approach the global minimum[1-3,6,10,25]. At $T_{pe}$ the gelation process ends; the result is a pseudogel with an inhomogeneous structure, comprising domains of various length scales. The other rate processes after $T_{pe}$ might correspond to the relaxation dynamics of water released from these various size domains during the syneresis process, leading to the continuous evolution of the internal structure of agar gels with the slowly relaxing breathing modes, as recently suggested by S. Boral et al[16].

More interestingly, the temperature dependence of $W_n$ in Fig.4c clearly indicates the existence of two Arrhenius-like relaxation processes in Stages II and III, which is quite consistent with the previously-reported HMP/sucrose gelation process[25]. Similar evidence also exists in the temperature dependence of the maximum SDR (i.e. $K_n$) as shown in Fig.4b. The crossover temperature of the two processes corresponds to the pseudo-equilibrium temperature $T_{pe}$. Motivated by previous Arrhenius-Equation treatments of temperature-dependent relaxation behaviors in HMP/sucrose gel[25] and polycrystalline metals[42-43], we used an empirical Arrhenius-like equation[43] $W_n \propto A_0 exp(-E_a/RT)$ (at which $A_0$ is the pre-exponential factor, $E_a$ denotes the apparent activation energy, and $R$ is the universal gas constant) to fit the $W_n \sim T_n^{-1}$ data from Stages II and III in Fig.4c. The resultant $E_{aII}$ and $E_{aIII}$ values are given in Table I. Although the fitting equation used above, $W_n \propto A_0 exp(-E_a/RT)$, is only an empirical equation and the parameter $E_a$ is not precisely equal to the activation energy of a Debye relaxation process described by the Arrhenius equation[b], we propose that the $W_n \sim T_n^{-1}$ result shown in Fig.4c and 4d and the derived $E_{aII}$ and $E_{aIII}$ values have true physical meaning.

From Table I we can see that the $E_{aII}$ value for $C=0.75-2.0\%$ w/w is in the range of 39~74KJ/mol, which is significantly larger than corresponding $E_{aIII}$ value of ~7.0 KJ/mol; the $E_{aII}$ value is also evidently dependent on the agar concentration. According to previous studies[3-8,14-20,22-26,34-39,44], it is reasonable to argue that the obtained $E_{aII}$ value of 39-74 KJ/mol may be phenomenologically ascribed to the energy barrier related to the formation of the agar pseudogel network during the agar gelation process in Stage II. In this process, the formation and fission of hydrogen bonding interactions between disconnected agar chains, fibers and crosslinking junctions of the agar gel structure play an important role. Increasing the agar concentration would increase the hydrogen bonding interaction, thus leading to an increase of the $E_{aII}$ value. The lower activation energy value (i.e. $E_{aIII}$~7.0 KJ/mol) is quite close to that for the viscous flow of water[45]. This similarity implies that the relaxation behavior in the agar pseudogel syneresis process (in Stage III) might be associated with only the contraction in volume of the agar gel and the consequent release of squeezed water, rather than the fission of the hydrogen bonds or rupture of the gel structure.

---

[b] The relaxation behavior of agar-water systems is more complicated than the Debye relaxation theory description[1-3,42-43].

## 4. Conclusion

In the present study, the kinetic behavior of the structural formation process in the agar-water gel system during natural cooling from ~80°C to room temperature was investigated by a time-resolved measurement based on our custom built torsion resonator. We find that the agar sol-gel transition during natural cooling is a gradual and complicated process, quite different from the nucleation and crystallization of the typical water-ice transition[29e,32-33] accompanied by abrupt changes in temperature and response strain (or $tg\phi$). From the time (or temperature) evolution of SDR, the agar natural-cooling gelation process can be clearly divided into three stages: the sol stage (Stage I), gelation zone (Stage II) and gel stage (Stage III). In Stages II and III, we also find that there exists a fluctuant decaying behavior of SDR, following the rapid increase early in Stage II. Such behavior could be interpreted as the sum of multiple relaxation processes, each having a SDR with different magnitude and different temperature dependence, and could be well described by a multiple-order Gaussian equation: $dG'/dt \equiv \sum_{n=0}^{m} dG'/dt^{(n)} = \sum_{n=0}^{m} K_n \exp[-2(T-T_n)^2/W_n^2]$. In this interpretation, the dispersions with the characteristic temperature $T_n$ (larger than $T_{pe}$) are attributed to the relaxation of the self-assembly of agar molecules, or the growth of junction zones en route to gelation in stages I and II. Those with smaller SDR magnitude (i.e. $K_n$) after $T_{pe}$ are attributed to the relaxation dynamics of water released from domains of various sizes during the syneresis process.

More interestingly, we find that the temperature dependence of $W_n$, before and after $T_{pe}$, follows an empirical Arrhenius-like law with different activation energies of $E_{aII}$ and $E_{aIII}$, respectively. The $E_{aII}$ value was predicted to lie in the range of 39-74 KJ/mol, which may be ascribed to the formation or fission of hydrogen bonding of the agar gel structure. In contrast, the low activation energy, $E_{aIII}\sim 7.0$ KJ/mol, may be ascribed to the release of water from agar hydrogels (called syneresis).

Finally, it should be emphasized that agar exhibits a more complex gelation mechanism than described here; this paper does not answer all of the questions related to the origin and mechanism of the structural formation process of such complex systems, but it makes an attempt to present some experimental results to provide a new insight into the kinetic behavior of the structural formation process in the agar-water gel system on natural cooling. We should also keep in mind that there are other interesting questions yet to be completely answered. For example, the multiple-order Gaussian equation (Eq.1) and the Arrhenius-like equation ($W_n \propto A_0 exp(-E_a/RT)$) are used as the empirical equations in this work, but their exact mathematical and physical mechanism still remains ambiguous. Additionally, this work focuses only on the case of the natural cooling, and the effect of the cooling rate needs further experimental studies in the future.

## Acknowledgements

This work was supported by the national physics base at Sun Yat-Sen University (Nos J0630320 and J0730313).

[41] Note that the *m* value depended on the detecting time (*t*) of the agar gelation process and the agar concentration, and might be dependent on working frequency of the measuring torsion resonator, the final cooling temperature in experiments, as well as on the cooling rate. Further experimental studies on the physical mechanism of Eq.1 will be reported in the near future, but the case of three agar-water solutions naturally cooled to room temperature were only concerned in this work.

**Figure Captions**

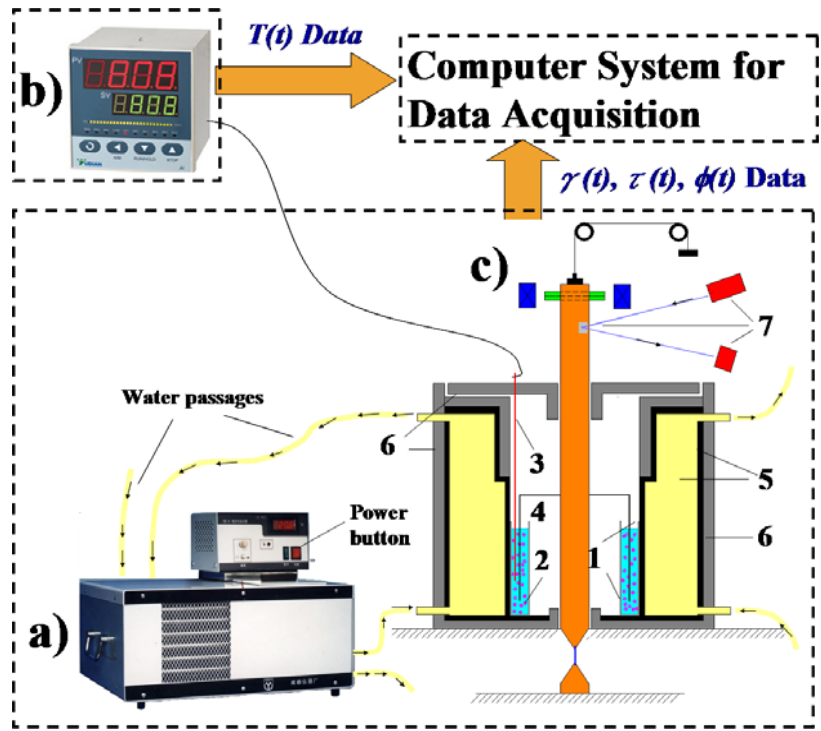

Fig.1. Schematic diagram of measuring system. a) HS-4 standard thermostatic water bath[31a]. b) High-precision *Yudian UGU* microprocessor temperature controller[31b]. c) Simplified diagram of the torsion resonator[29]: 1. The Teflon sample cell of annular-cup type (with inner diameter of 10 mm, outer diameter of 18 mm and wall-thickness of 0.5 mm); 2. measured sample; 3. High-precision thermocouple; 4. Inverted cup (with wall-thickness of ~0.3mm); 5. Ethylene glycol-water mixture and copper temperature housing, 6. Teflon jacket; 7. Optical measuring system[29a].) For more details of torsion resonator see our recent references[29a-c].

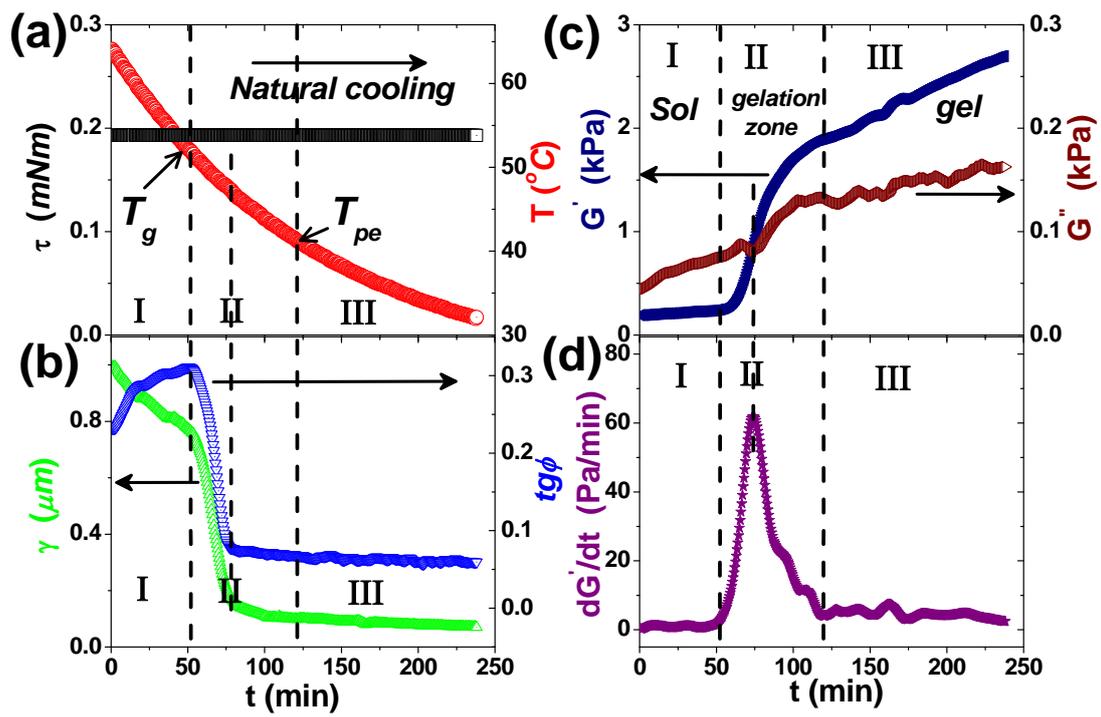

Fig.2. Gelation kinetics of an agar aqueous solution (2% w/w) upon natural cooling from about 70°C to 30°C.

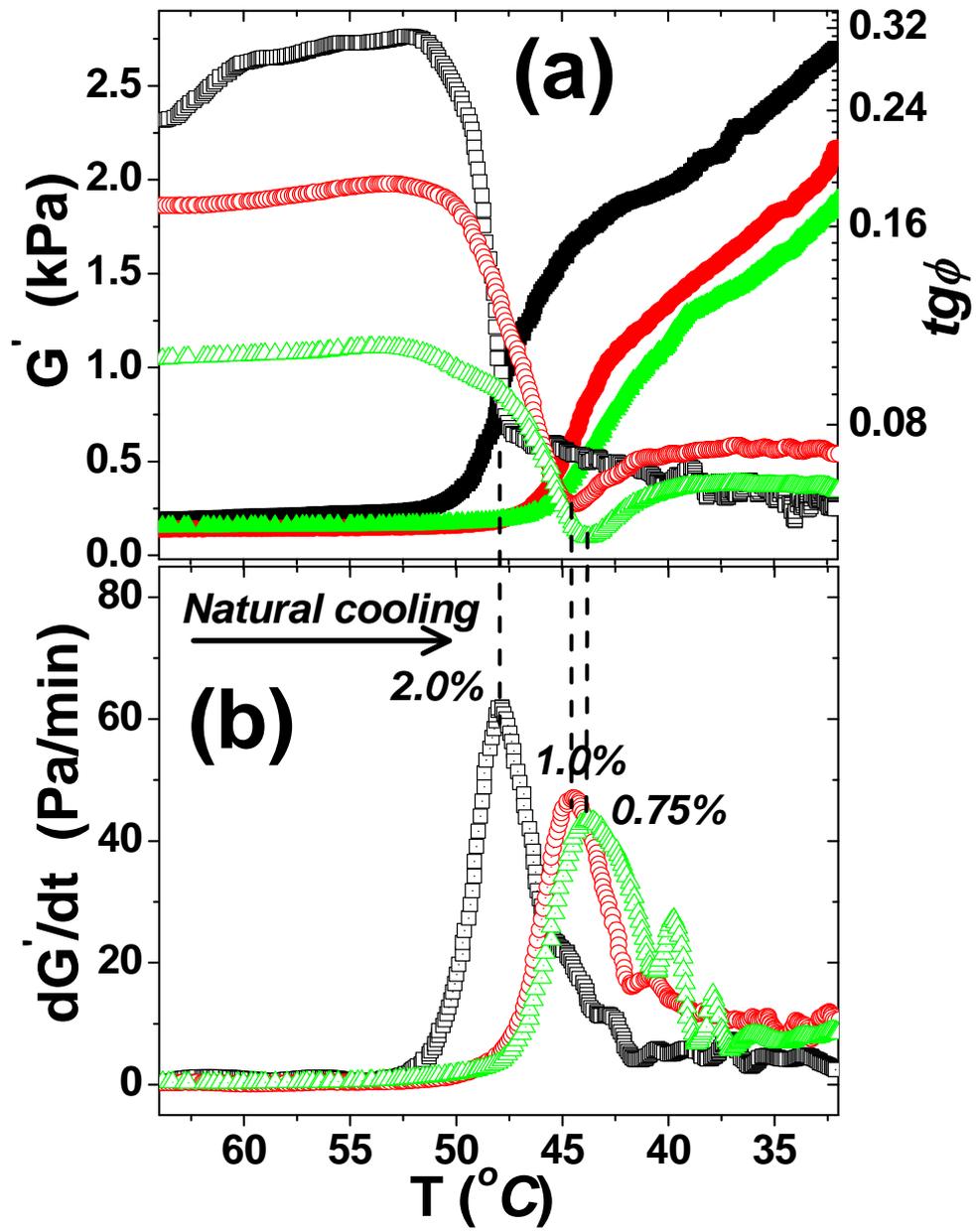

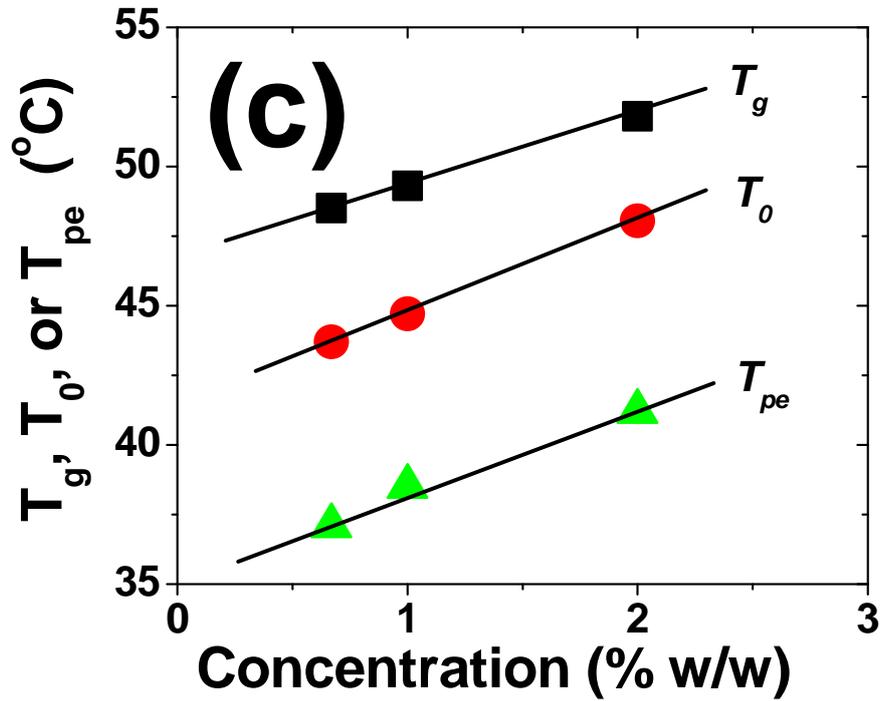

Fig.3. Temperature dependences of (a) the storage modulus, $G'$, and the loss, $tg\phi \equiv G''/G'$, as well as (b) the gelation rate, $dG'/dt$, for agar-water solutions at concentrations of 2.0%, 1.5% and 0.75% w/w upon natural cooling. (c) The concentration dependence of values of the temperature parameters, $T_g$, $T_0$, and $T_{pe}$, which were obtained by using the multiple-order Gaussian equation to fit the $dG'/dt \sim T$ data of Fig.3b (Typical fitting process for the case of $C=2.0\%$ w/w is given in Fig.4a).

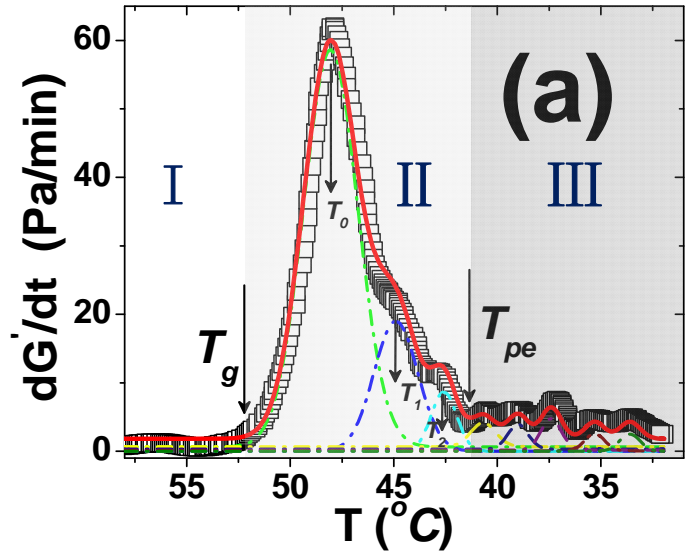

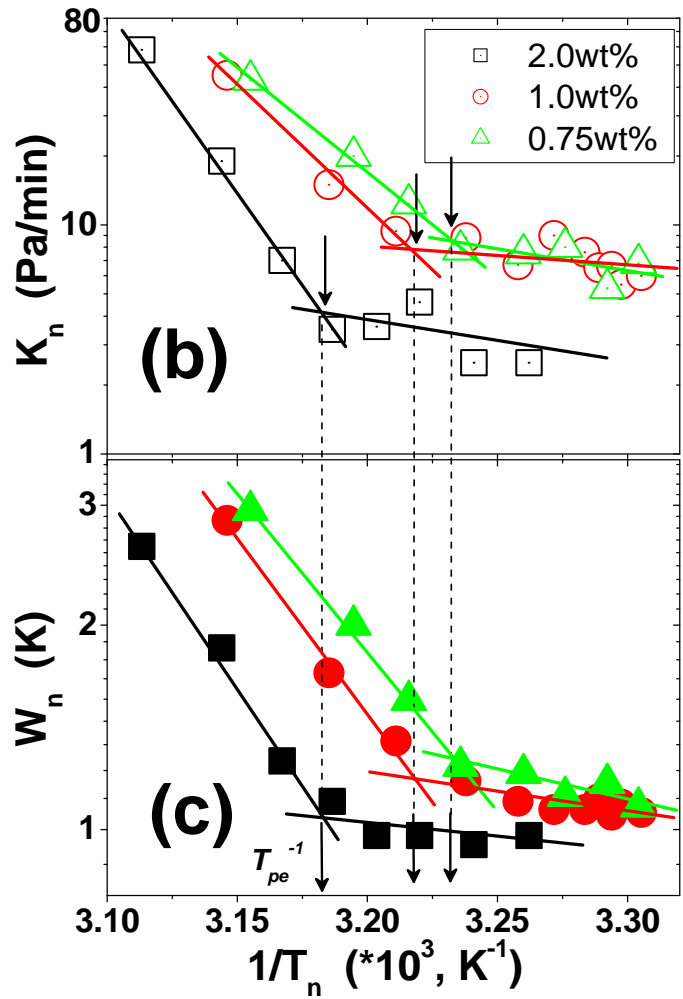

Fig.4. (a) Temperature dependence of the gelation rate, characterized by $dG'/dt$, for the natural cooling process of the *2% w/w* aqueous agarose solution. Open symbols show the experimental results; several different dashed curves (exhibiting various color online) are fitted by the Gaussian equation, $y_n(T)=K_n*exp[-2(T-T_n)^2/W_n^2]$, with three fitting parameters ($K_n$, $T_n$, and $W_n$); the solid curve is the sum of these individual Gaussian fitting results (the multiple-order Gaussian equation, $\frac{dG'}{dt}=\sum_{n=0}^{m}y_n(T)$, where *m=7* used for the case of *C=2% w/w*). (b) and (c) show the Arrhenius treatment[25,42-43] for the effect of temperature to the maximal gelation rate ($K_n$) and $W_n$ of *C=0.75%, 1.0%* and *2% w/w* agar-water systems in Stages II and III.

## Tables

**Table I.** Apparent activation energies of agar-water systems in Stages II and III.

| Agar Concentration (% w/w) | $E_{aII}$ (KJ/mol) | $E_{aIII}$ (KJ/mol) |
|---|---|---|
| 2.0 | 73.7 | 6.7 |
| 1.0 | 52.1 | 7.1 |
| 0.75 | 39.3 | 7.7 |